\documentclass[aps]{revtex4}%
\usepackage{amsfonts}
\usepackage{amsmath}
\usepackage{amssymb}
\usepackage{graphicx}%
\providecommand{\U}[1]{\protect\rule{.1in}{.1in}}
\begin{document}
\title{Relaxation- and Decoherence-free subspaces in networks of weakly and strongly
coupled resonators}
\author{M. A. de Ponte, S. S. Mizrahi, and M. H. Y. Moussa}
\affiliation{Departamento de F\'{\i}sica, Universidade Federal de S\~{a}o Carlos, Rodovia
Washington Luiz km 235, S\~{a}o Carlos, 13565-905, SP, Brazil}

\begin{abstract}
We consider a network of interacting resonators and analyze the physical
ingredients that enable the emergence of relaxation-free and decoherence-free
subspaces. We investigate two different situations: i) when the whole network
interacts with a common reservoir and ii) when each resonator, strongly
coupled to each other, interacts with its own reservoir. Our main result is
that both subspaces are generated when all the resonators couple with the same
group of reservoir modes, thus building up a correlation (among these modes),
which has the potential to shield particular network states against relaxation
and/or decoherence.

\end{abstract}
\date{\today}

\pacs{PACS numbers: 32.80.-t, 42.50.Ct, 42.50.Dv}
\maketitle

The search for mechanisms to bypass decoherence, a subject of major concern
for quantum information processing, has deepened our understanding of open
quantum systems and led to ingenious schemes for coherence control, which go
far beyond the quest for conditions that weaken the system-reservoir coupling
\cite{Landauer,Unruh}. The myriad contributions to this subject started,
inspired by classical error-correcting codes, with quantum coding schemes for
information stored in a quantum memory \cite{QECC}. On the assumption that the
decoherence process acts independently on each of the qubits stored in a
memory, a particular qubit encoded in a block of ancillary qubits is able to
withstand a substantial degree of interaction with the reservoir without
degradation of its information. Another strategy, the so-called engineering
reservoirs \cite{Poyatos}, compels the system of interest, whose state is to
be protected against decoherence, to engage in additional interactions besides
that with the reservoir. This program, based on the indirect control of
system-reservoir dynamics, has been developed for trapped ions
\cite{King,Matos} and atomic two-level systems \cite{Lutkenhaus,Agarwal} as
well. Finally, the process of collective decoherence, where a composite system
interacts with a common reservoir, has also instigated several interesting
results related to what has been called a decoherence-free subspace (DFS)
\cite{ZR,DFS,Lidar}. It is noteworthy that while quantum error-correcting
codes presuppose quantum systems that decohere independently, DFS -- as it has
been understood until the present study -- is generated by distinct quantum
systems coupled to a common reservoir.

In this contribution we are concerned with collective dissipation and
decoherence in a network of $N$\ coupled resonators. We analyze the physical
ingredients ruling the emergence of a DFS and, in particular, a
relaxation-free subspace (RFS) composed of states protected against both
dissipation and decoherence, demonstrating that a DFS contains a RFS. We first
analyze the situation, treated in the literature to date, where all the
resonators are coupled to a common reservoir. However, as the scenario of a
common reservoir is in practice rather unusual, we next analyze the situation
where each resonator interacts with its own reservoir, which seems to be more
appropriate for most physical systems. In the domain of cavity quantum
electrodynamics, distinct reservoirs must be considered for distinguishable
cavities, even if they present equal quality factors, inasmuch as there are no
correlations, whatsoever, between the reservoirs. The same applies to
distinguishable trapped ions or a traveling field reaching distinguishable
optical elements. There are a few particular situations where a set of quantum
systems do interact with a common reservoir, such as an atomic sample or
distinct fields inside a perfect cavity. In the former case, different atomic
transitions couple with different reservoir modes (RM) and again, the
correlations between these RM define either a common or distinct reservoirs,
as will be demonstrated below. In the latter, we also show that the proximity
of the distinct field modes sets the strength of the correlation function
between the RM which governs the emergence of both RFS and DFS.

\textit{A common reservoir}. Consider the Hamiltonian $H=H_{S}+H_{R}+H_{I}$,
for the system of $N$ interacting resonators ($H_{S}$), the reservoir ($H_{R}
$), and the interaction between the resonators and the reservoir ($H_{I}$).
Assuming, from here on, that the subscripts $m$ ($m^{\prime}$) and $n$
($n^{\prime}$), labelling the resonators, run from $1$ to $N$, the Hamiltonian
is given by ($\hbar=1$)
\begin{subequations}
\begin{align}
H_{S}  &  =\sum\nolimits_{m}\omega_{m}a_{m}^{\dagger}a_{m}+\frac{1}{2}%
\sum\nolimits_{m,n(m\neq n)}\lambda_{mn}\left(  a_{m}^{\dagger}a_{n}%
+\mathrm{h{.c.}}\right)  \mathrm{{,}}\label{1a}\\
H_{R}  &  =\sum\nolimits_{k}\widetilde{\omega}_{k}b_{k}^{\dagger}%
b_{k}\mathrm{{,\ \ }}H_{I}=\sum\nolimits_{m,k}V_{mk}\left(  a_{m}%
b_{k}^{\dagger}+\mathrm{h{.c.}}\right)  \mathrm{{,}} \label{1b}%
\end{align}
where $a_{m}^{\dagger}$ ($a_{m}$) is the creation (annihilation) operator of
mode $\omega_{m}$, while $b_{k}$ and $b_{k}^{\dagger}$ are the analogous
operators of the $k$th bath mode, whose corresponding frequency and coupling
strength to the $m$th resonator are $\widetilde{\omega}_{k}$ and $V_{mk}$,
respectively. We assume that the interactions between the resonators of
frequencies $\omega_{m}$ and $\omega_{n}$ are within the rotating wave
approximation with coupling strength $\lambda_{mn}$. In what follows we
analyze both the weak and strong coupling regimes between the resonators,
where $N\lambda_{mn}\ll\omega_{m\prime}$ and $N\lambda_{mn}\gtrsim
\omega_{m\prime}$, respectively \cite{Mickel}. We also investigate the special
case of the weak coupling regime, of negligible mutual interaction between the
resonators, where $\lambda_{mn}\ll\left(  \omega_{m^{\prime}}-\omega
_{n^{\prime}}\right)  $.

The Hamiltonian $H_{S}$ can be brought into a diagonal form $H_{S}=\sum
_{m}\Omega_{m}A_{m}^{\dagger}A_{m}$ through a linear transformation
$A_{m}=\sum_{n}C_{mn}a_{n}$, $\left\{  \Omega_{m}\right\}  $ being the normal
modes of the network and $C_{mn}$ the elements of an orthogonal and symmetric
matrix ($C_{mn}=C_{nm}=C_{mn}^{-1}$). The interaction goes into the form
$H_{I}=\sum_{m,n,k}C_{mn}V_{mk}\left(  A_{n}b_{k}^{\dagger}+\mathrm{h{.c.}%
}\right)  $. Therefore, instead of $N$ coupled resonators interacting with a
common reservoir, as in Eqs. (\ref{1a}) and (\ref{1b}), the transformed
Hamiltonian describes $N$ decoupled resonators under the action of the
reservoir. In the picture defined by the transformation $\operatorname*{e}%
\nolimits^{-i\left(  H_{S}+H_{R}\right)  t}$, we obtain the interaction
$\mathbf{V}(t)=\sum_{m,n}\left(  \mathcal{O}_{mn}^{\dagger}(t)A_{n}%
+\mathrm{h{.c.}}\right)  $, where $\mathcal{O}_{mn}^{\dagger}(t)=\sum
_{k}C_{mn}V_{mk}b_{k}^{\dagger}\operatorname*{e}\nolimits^{i\left(
\widetilde{\omega}_{k}-\Omega_{n}\right)  t}$. By analogy with the
developments in Refs. \cite{Mickel} and noting that $C_{mn}<1$, we obtain, at
$T=0$K, the master equation of the reduced system of resonators
\end{subequations}
\begin{equation}
\frac{d\rho_{1,\ldots,N}(t)}{dt}=i[\rho_{1,\ldots,N}(t),H_{S}]+\sum
\nolimits_{m,n}\mathcal{L}_{mn}(t)\rho_{1,\ldots,N}(t)\mathrm{{,}} \label{3}%
\end{equation}
where the influence of the reservoirs is represented by the Liouville
operators $\mathcal{L}_{mn}\bullet$, accounting for direct ($m=n$) and
cross-decay ($m\neq n$) channels \cite{Mickel}, which link together the many
resonators via a common reservoir. The Liouville operators are given by%
\begin{equation}
\mathcal{L}_{mn}(t)\rho_{1,\ldots,N}(t)=\frac{\gamma_{mn}}{2}\left\{  \left[
a_{n}\rho_{1,\ldots,N}(t),a_{m}^{\dagger}\right]  +\mathrm{h{.c.}}\right\}  .
\label{5}%
\end{equation}
where the coefficients $\gamma_{mn}=\sum_{m^{\prime},n^{\prime}}%
\varepsilon_{mm^{\prime}}\left(  \Omega_{n^{\prime}}\right)  C_{m^{\prime
}n^{\prime}}C_{n^{\prime}n}$, take into account the elements of the
transformed matrix, coming from the topology of the network: the frequency
distribution of the resonators, their coupling pattern and strengths.
$\gamma_{mn}$ also depends on the interaction between the system and the
reservoir through the function $\varepsilon_{mn}\left(  \Omega_{m^{\prime}%
}\right)  \equiv\int_{0}^{t}dt^{\prime}\int_{0}^{\infty}\frac{d\widetilde
{\omega}}{\pi}\sigma\left(  \widetilde{\omega}\right)  V_{m}\left(
\widetilde{\omega}\right)  V_{n}\left(  \widetilde{\omega}\right)
\operatorname*{e}\nolimits^{-i\left(  \widetilde{\omega}-\Omega_{m^{\prime}%
}\right)  \left(  t-t^{\prime}\right)  }$, where the frequencies of the
reservoir are assumed to be continuous variable with spectral density
$\sigma(\widetilde{\omega}) $. The function $\varepsilon_{mn}\left(
\Omega_{m^{\prime}}\right)  $ is related to the correlation between the RM
\cite{Mickel}, through $\int_{0}^{t}dt^{\prime}\left\langle \mathcal{O}%
_{mm^{\prime}}(t)\mathcal{O}_{nn^{\prime}}^{\dagger}(t^{\prime})\right\rangle
\approx\frac{1}{2}\varepsilon_{mn}\left(  \Omega_{n^{\prime}}\right)
\operatorname*{e}\nolimits^{-i\left(  \Omega_{n^{\prime}}-\Omega_{m^{\prime}%
}\right)  t}${.}

To analyze the behavior of the correlation $\varepsilon_{mn}\left(
\Omega_{m^{\prime}}\right)  $ we first assume, as usual, that both the
spectral density $\sigma\left(  \widetilde{\omega}\right)  $ and the coupling
strength $V_{m}(\widetilde{\omega})$ are functions that vary slowly around the
frequency $\Omega_{m}$, such that $\varepsilon_{mn}\left(  \Omega_{m^{\prime}%
}\right)  \approx\sigma\left(  \Omega_{m^{\prime}}\right)  V_{m}\left(
\Omega_{m^{\prime}}\right)  V_{n}\left(  \Omega_{m^{\prime}}\right)
$.\textbf{\ }The system-reservoir interaction, which applies to all coupling
regimes, is thus modeled by the function \cite{Mickel}
\begin{equation}
V_{m}\left(  \omega\right)  =V_{m}\left(  \sum\nolimits_{j=1}^{M}%
\operatorname*{e}\nolimits^{-\xi_{j}\left(  \widetilde{\omega}-\Omega
_{j}\right)  ^{2}}\right)  ^{1/2}\mathrm{{.}} \label{8}%
\end{equation}
We observe from Eq. (\ref{8}) that when the separation between two normal
modes is large enough compared to the Gaussian widths,\textbf{\ }$\left|
\Omega_{j}-\Omega_{j^{\prime}}\right|  \gg$ $\xi_{j}^{-1}$,$\xi_{j^{\prime}%
}^{-1}$, these modes are considered distinct. The difference between the weak
and strong coupling regime is that, in the former, the normal modes
$\Omega_{m}$ are approximately equal to the natural frequencies $\omega_{m}$,
while in the latter, they may be quite different, depending on the coupling
strength $\lambda_{mn}$ \cite{Mickel} (see below). In both cases the
interactions between the resonators couple them with all groups of bath modes
around\textbf{\ }all the $M\left(  \leq N\right)  $\ distinct normal modes
$\Omega_{j}$\ (with $j=$\ $1,...,M$) defined by the topology of the network
\cite{Mickel}. In fact, it has been demonstrated \cite{Mickel} that the states
of two distinct resonators, say $m$ and $n$, are interchanged many times
before the relaxation takes place, in both weak and strong coupling regimes,
the recurrence time of the states being approximately $\left(  N\lambda
_{mn}\right)  ^{-1}$. This interchange of excitation is the mechanism behind
the coupling of each resonator to all groups of RM around all normal modes.
When considering resonators with negligible mutual interaction, such that the
$m$th resonator practically remains at its natural frequency $\omega_{m}$,
expression (\ref{8}) degenerates to the expected result $V_{m}\left(
\widetilde{\omega}\right)  =V_{m}\operatorname*{e}\nolimits^{-\xi_{m}\left(
\widetilde{\omega}-\omega_{m}\right)  ^{2}/2}$, where the $m$th resonator
couples only with the group of RM around $\omega_{m}$. In fact, for negligible
mutual interaction the cross-decay channels are exceedingly small, around
$\lambda_{mn}/\left(  \omega_{m^{\prime}}-\omega_{n^{\prime}}\right)  $,
leading to recurrence times longer than the relaxation of the network.

Under the above considerations, for both weak and strong coupling regimes, we
obtain the result%
\begin{equation}
\varepsilon_{mn}\left(  \Omega_{\ell}\right)  \approx\sigma\left(
\Omega_{\ell}\right)  V_{m}V_{n}\left(  \sum\nolimits_{j,j^{\prime}%
}\operatorname*{e}\nolimits^{-\xi_{j}\left(  \Omega_{\ell}-\Omega_{j}\right)
^{2}-\xi_{j^{\prime}}\left(  \Omega_{\ell}-\Omega_{j^{\prime}}\right)  ^{2}%
}\right)  ^{1/2} \label{9}%
\end{equation}
revealing that $\varepsilon_{mn}\left(  \Omega_{m^{\prime}}\right)
\approx\sigma\left(  \Omega_{m^{\prime}}\right)  \left|  V_{m}V_{n}\right|
\equiv\Gamma$ whenever $\left|  V_{n}\right|  \approx\left|  V_{m}\right|
\equiv\left|  V\right|  $ and $\sigma\left(  \Omega_{m}\right)  $ is a
constant $\sigma$, which is a reasonable assumption for a Markovian white
noise, where $\Gamma\equiv\sigma\left|  V\right|  ^{2}$. Note that only the
term $j=j^{\prime}=\ell$ contributes significantly to each sum in Eq.
(\ref{9}), in both weak and strong coupling regimes. The result $\varepsilon
_{mn}\left(  \Omega_{m^{\prime}}\right)  \approx\Gamma$ implies that
$\gamma_{mn}=\Gamma$ and, consequently, both direct and cross-decay channels
have the same order of magnitude, a crucial condition, as demonstrated below,
for particular classes of states to be shielded against relaxation and/or
decoherence. In the case of negligible mutual interactions between the
resonators, we obtain $\varepsilon_{mn}\left(  \omega_{m^{\prime}}\right)
\approx\sigma V_{m}V_{n}\operatorname*{e}\nolimits^{-\left[  \xi_{m}\left(
\omega_{m^{\prime}}-\omega_{m}\right)  ^{2}+\xi_{n}\left(  \omega_{m^{\prime}%
}-\omega_{n}\right)  ^{2}\right]  /2}\mathrm{{,}}$ which, differently from Eq.
(\ref{9}), varies from $0$ to $\Gamma$, depending on the proximity between the
natural frequencies $\omega_{m}$. The correlation function $\varepsilon
_{mn}\left(  \omega_{m^{\prime}}\right)  $ attains the maximum value $\Gamma$,
bringing about a DFS, only in the degenerate case where all the resonators
have the same natural frequency. We stress that the maximum correlation
function ($\varepsilon_{mn}=\Gamma$) between the RM will be induced by the
network\ itself when all its resonators interact with the same groups of RM
--- those around the degenerate frequency in the case of negligible mutual interactions.

In order to characterize the RFS and DFS, from here on we consider the
degenerate case where $\omega_{m}=\omega$ and $\lambda_{mn}=\lambda$, leading
to two distinct normal modes $\Omega_{+}=$ $\omega+\left(  N-1\right)
\lambda$ and $\Omega_{-}=\omega-\lambda$ \cite{Mickel}. \ We also consider a
Markovian white noise reservoir, where $\varepsilon_{mn}$ $\approx\varepsilon$
and $\Gamma_{m}\approx\Gamma$, and the initial entanglement between the $N$
resonators is given by $\left\vert \Psi_{1,\ldots,N}\right\rangle
=\mathcal{N}\sum_{r=1}^{J}\Upsilon_{r}\left\vert \left\{  \beta_{m}%
^{r}\right\}  \right\rangle $. This collective state comprehends a
superposition of $J$ product states, each one consisting of a product of $N$
distinct coherent states $\beta_{m}^{r}$, $\mathcal{N}$ being a normalization
factor . We also redefine the correlation function $\varepsilon$, introducing
the dimensionless parameter $\epsilon=\varepsilon/\Gamma$ varying from zero to
unity. Under these assumptions and following the reasoning in Ref.
\cite{Mickel}, we obtain the density operator for both regimes $\epsilon
\cong1$\ and $\epsilon\ll1$:%
\begin{equation}
\rho_{1,...,N}(t)=\mathcal{N}^{2}\sum\nolimits_{r,s}\Upsilon_{r}\Upsilon
_{s}^{\ast}\frac{\left\langle \left\{  \beta_{m}^{r}\right\}  \right.
\left\vert \left\{  \beta_{m}^{s}\right\}  \right\rangle }{\left\langle
\left\{  \zeta_{m}^{r}\right\}  \left\vert \left\{  \zeta_{m}^{s}\right\}
\right.  \right\rangle }\left\vert \left\{  \zeta_{m}^{s}\right\}
\right\rangle \left\langle \left\{  \zeta_{m}^{r}\right\}  \right\vert ,
\label{12}%
\end{equation}
where $\zeta_{m}^{r}=\sum_{n}\Theta_{mn}(t)\beta_{n}^{r}$ and the function
accounting for the excitation decay is given by $\Theta_{mn}(t)=\frac
{\operatorname*{e}\nolimits^{-(1-\epsilon)\Gamma t/2}}{N}\left(
\operatorname*{e}\nolimits^{-(\epsilon N\Gamma/2+i\Omega_{+})t}+\left(
N\delta_{mn}-1\right)  \operatorname*{e}\nolimits^{-i\Omega_{-}t}\right)  $.

\textit{Decoherence Free Subspace}. To analyze the role played by the
excitation decay function $\Theta_{mn}(t)$ in the emergence of a DFS, let us
consider a particular family of initial states $\left|  \Psi_{1,\ldots
,N}\right\rangle $ whose representative element is
\begin{equation}
\left|  \Psi_{1,...,N}\right\rangle _{R,S}=\mathcal{N}_{\pm}\left(  \left|
\underset{R}{\underbrace{\alpha,\ldots,\alpha}},\underset{S}{\underbrace
{-\alpha,\ldots,-\alpha}}\right\rangle \pm\left|  \underset{R}{\underbrace
{-\alpha,\ldots,-\alpha}}\underset{S}{,\underbrace{\alpha,\ldots,\alpha}%
}\right\rangle \right)  \otimes\left|  \underset{N-R-S}{\underbrace
{\eta,\ldots,\eta}}\right\rangle \mathrm{,} \label{13}%
\end{equation}
where $R$ ($S$) indicates the number of resonators in the coherent state
$\alpha$ ($-\alpha$) in the first term of the superposition and $-\alpha$
($\alpha$) in its second term. The remaining $N-R-S$ resonators are in the
coherent states $\left\{  \eta\right\}  $. As we are considering a symmetric
network where all the resonators have the same damping rate, they are
indistinguishable. Therefore, swapping the states of any two resonators $m$
and $n$ we obtain a state which is completely equivalent to Eq. (\ref{13}).
For the states defined by Eq. (\ref{13}) and the evolution (\ref{12}), we
obtain
\begin{equation}
\zeta_{m}^{r}=\sum\nolimits_{n}\frac{\beta_{n}^{r}}{N}\operatorname*{e}%
\nolimits^{-\left[  1+\left(  N-1\right)  \epsilon\right]  \Gamma
t/2-i\Omega_{+}t}+\left(  \beta_{m}^{r}-\sum\nolimits_{n}\frac{\beta_{n}^{r}%
}{N}\right)  \operatorname*{e}\nolimits^{-\left(  1-\epsilon\right)  \Gamma
t/2-i\Omega_{-}t}\mathrm{{,}} \label{14}%
\end{equation}
from which two distinct classes of states follow: the superposition states
$\left|  \Psi_{1,...,N}\right\rangle _{R=S=N/2}$ with an effective lower decay
rate\textbf{\ }(compared to $\Gamma$) $\Gamma_{\downarrow}=\left(
1-\epsilon\right)  \Gamma$, since $\zeta_{m}^{r}=\beta_{m}^{r}%
\operatorname*{e}\nolimits^{-\Gamma_{\downarrow}t/2-i\Omega_{-}t}$, and the
product states $\left|  \Psi_{1,...,N}\right\rangle _{R=S=0}$ with an enhanced
decay rate $\Gamma_{\uparrow}=$ $\left[  1+\left(  N-1\right)  \epsilon
\right]  \Gamma$, since $\zeta_{m}^{r}=\eta\operatorname*{e}\nolimits^{-\Gamma
_{\uparrow}t/2-i\Omega_{+}t}$. For the decoherence time of the state
(\ref{13}) we obtain
\begin{equation}
\tau_{D}\left(  \left|  \Psi_{1,\ldots,N}\right\rangle \right)  =\frac
{1}{2\left|  \alpha\right|  ^{2}}\frac{1}{\left(  R-S\right)  ^{2}%
\Gamma+\left[  \left(  R+S\right)  -\left(  R-S\right)  ^{2}\right]
\Gamma_{\downarrow}}\mathrm{{,}} \label{15}%
\end{equation}
to be analyzed for both classes of states $\left|  \Psi_{1,...,N}\right\rangle
_{R=S=N/2}$ and $\left|  \Psi_{1,...,N}\right\rangle _{R=S=0}$, under both
regimes $\epsilon\cong1$ and $\epsilon\ll1$. In the regime $\epsilon\cong1$,
such that $\Gamma_{\downarrow}\cong0$ and $\Gamma_{\uparrow}\cong N\Gamma$,
both classes of states belong to a larger DFS defined by the condition $R=S$.
However, while the superpositions $\left|  \Psi_{1,...,N}\right\rangle
_{R=S=N/2}$, with $\Gamma_{\downarrow}\cong0$, are effectively uncoupled from
the reservoir, defining a RFS, it is worth noting that the states $\left|
\Psi_{1,...,N}\right\rangle _{R=S=0} $ dissipate at the maximum rate
$\Gamma_{\uparrow}\cong N\Gamma$. The intermediate class of decoherence-free
states, where $R=S<N/2$, also dissipate at the maximum rate $\Gamma_{\uparrow
}\cong N\Gamma$, leading to the conclusion that the DFS contains the RFS. The
correlation $\epsilon\cong1$ is achieved independently of the spectral shape
of the network for both weak and strong coupling regimes; in the case of
negligible interaction between the resonators the regime $\epsilon\cong1$
occurs only when all the resonators have approximately the same natural
frequency. These three situations, where the resonators interact with the same
group of RM, trigger a kind of phase matching, between the resonators
themselves together with the RM, which is unequivocally produced by the
cross-decay channels, as discussed below.

For the class of states not belonging to the DFS, i.e., with $R\neq S$, as for
example $\left|  \Psi_{1,...,N}\right\rangle _{R=1,S=0}=N_{\pm}\left(  \left|
\alpha\right\rangle \pm\left|  -\alpha\right\rangle \right)  _{1}%
\otimes\left|  \left\{  \eta_{j}\right\}  \right\rangle $\textbf{\ }(where a
Schr\"{o}dinger-cat like state is prepared in resonator $1$ while all the
remaining $j$ resonators are prepared in the coherent states $\eta$), we get
the expected decoherence time $\tau_{D}\left(  \left|  \Psi_{1,...,N}%
\right\rangle _{R=1,S=0}\right)  =\Gamma^{-1}/(2\left|  \alpha_{1}\right|
^{2})$, even when $\epsilon\cong1$. Finally, in the regime $\epsilon\ll1$,
only the product states $\left|  \Psi_{1,...,N}\right\rangle _{R,S=0}$ remain
in the DFS, the decoherence time for the superpositions $\left|
\Psi_{1,...,N}\right\rangle _{R=S=N/2}$ increasing as the correlation
$\epsilon$ decreases.

It is worth noting that the necessary and sufficient condition for a subspace
to be decoherence-free can be established through the Lindblad master
equation
\begin{equation}
\frac{d\rho_{1,\ldots,N}(t)}{dt}=i[\rho_{1,\ldots,N}(t),H_{S}]+\frac{1}{2}%
\sum\nolimits_{m,n}\Gamma_{mn}\left(  \left[  \Lambda_{m}\rho_{1,\ldots
,N}(t),\Lambda_{n}^{\dagger}\right]  +\mathrm{h{.c.}}\right)  \mathrm{{,}}
\label{16}%
\end{equation}
in that all these subspace states $\left\{  \Psi\right\}  $ must be degenerate
eigenstates of all the Lindblad operators $\{\Lambda_{m}\}$, i.e.,
$\Lambda_{m}\left|  \Psi\right\rangle =c_{m}\left|  \Psi\right\rangle $,
$\forall$ $m,\Psi$ \cite{Lidar}. This is exactly what happens to Eq. (\ref{3})
when considering, for a reservoir at $T=0$K, the above-mentioned three cases
where $\epsilon\cong1$. In fact, under this condition, Eq. (\ref{3}) can be
written as
\begin{equation}
\frac{d\rho_{1,\ldots,N}(t)}{dt}=i[\rho_{1,\ldots,N}(t),H_{S}]+\frac{N\Gamma
}{2}\left(  \left[  \Lambda\rho_{1,\ldots,N}(t),\Lambda^{\dagger}\right]
+\mathrm{h{.c.}}\right)  , \label{17}%
\end{equation}
with the eigenvalue equation $\Lambda\left|  \Psi_{1,...,N}\right\rangle
_{R=S}=\left(  \eta\left(  N-2R\right)  /\sqrt{N}\right)  \left|
\Psi_{1,\ldots,N}\right\rangle _{R=S}$ for the operator $\Lambda=\sum_{n}%
a_{n}/\sqrt{N}$. As far as $\Lambda\left|  \Psi_{1,...,N}\right\rangle
_{R=S=N/2}=0$ while $\Lambda\left|  \Psi_{1,...,N}\right\rangle _{R=S=0}%
=\eta\sqrt{N}\left|  \Psi_{1,\ldots,N}\right\rangle _{R=S=0}$, we conclude
that: while decoherence-free states are characterized by an eigenvalue
equation, dissipative-free states follow from the fact that their associated
eigenvalues are null \cite{ZR}. To summarize, we note that the crucial
ingredient for establishing a DFS is the action of cross-decay channels
$L_{mn}(t)\bullet$ which always take place for a common reservoir. When the
cross-decay channels are of the same order of magnitude as the direct-decay
channels $L_{mn}(t)\bullet$, such that $\epsilon\cong1$, a DFS emerges for the
states where $R=S$. When $R=S=N/2$, these states compose a RFS.

\textit{Strong coupling and distinct reservoirs}. To get round the necessity
of a common reservoir for the emergence of a DFS, we briefly analyze the case
where each resonator is coupled to a particular reservoir. Now, however, they
must be strongly coupled to each other for the accomplishment of the
conditions leading to a DFS. Considering the notation employed in Eqs.
(\ref{1a}) and (\ref{1b}), the system Hamiltonian in the degenerate case
$\omega_{m}=\omega_{0}$ and $\lambda_{mn}=\lambda$, is given by \cite{Mickel}
\begin{subequations}
\begin{align}
H_{S}  &  =\widetilde{\omega}_{0}\sum\nolimits_{m}a_{m}^{\dagger}a_{m}%
+\frac{\lambda}{2}\sum\nolimits_{m,n(m\neq n)}\left(  a_{m}^{\dagger}%
a_{n}+\mathrm{h{.c.}}\right)  \mathrm{{,}}\label{18a}\\
H_{R}  &  =\sum\nolimits_{m,k}\omega_{mk}b_{mk}^{\dagger}b_{mk}\mathrm{{,\ }%
}H_{I}=\sum\nolimits_{m,k}V_{mk}\left(  a_{m}^{\dagger}b_{mk}+\mathrm{h{.c.}%
}\right)  \mathrm{{,}} \label{18b}%
\end{align}
where $\widetilde{\omega}_{0}=\omega_{0}\left\{  1+\left[  (N-1)\left(
\lambda/2\omega_{0}\right)  \right]  ^{2}\right\}  $, under the assumption
that the energy spectrum has a lower bound \cite{Mickel}. As shown in Ref.
\cite{Mickel}, in the strong coupling limit $N\lambda\gtrsim\widetilde{\omega
}_{0}$ the master equation reads
\end{subequations}
\begin{equation}
\frac{d\rho_{1,...,N}(t)}{dt}=i\left[  \rho_{1,...,N}(t),H_{S}\right]
+\sum\nolimits_{m,n}\mathcal{L}_{mn}\rho_{1,...,N}(t)\mathrm{,} \label{19}%
\end{equation}
where $H_{S}=\widetilde{\omega}_{0}\sum_{m}a_{m}^{\dagger}a_{m}+(\lambda
/2)\sum_{m,n(m\neq n)}\left(  a_{m}^{\dagger}a_{n}+\mathrm{h{.c.}}\right)  $,
and the Liouville operators $\mathcal{L}_{m}\rho_{1,...,N}(t)$ are given by
\begin{equation}
\mathcal{L}_{mn}\rho_{1,...,N}(t)=\frac{\widetilde{\gamma}_{mn}}{2}\left\{
\left(  \left[  a_{n}\rho_{1,...,N}(t),a_{m}^{\dagger}\right]  +\mathrm{h{.c.}%
}\right)  \right\}  \mathrm{{.}} \label{21}%
\end{equation}
The coefficient $\widetilde{\gamma}_{mn}=\sum_{n^{\prime}}C_{mn^{\prime}}%
^{-1}\Gamma_{m}\left(  \Omega_{n^{\prime}}\right)  C_{n^{\prime}n}$ follows
from $\gamma_{mn}$ assuming $\varepsilon_{mm^{\prime}}\left(  \Omega
_{n^{\prime}}\right)  =\Gamma_{m}\left(  \Omega_{n^{\prime}}\right)
\delta_{mm^{\prime}}$. For a symmetric network, in the degenerate case
described by the Hamiltonian in Eqs. (\ref{18a}) and (\ref{18b}), only two
distinct normal modes arise: $\Omega_{+}=$ $\widetilde{\omega}_{0}+\left(
N-1\right)  \lambda$ and $\Omega_{-}=\widetilde{\omega}_{0}-\lambda$.
Consequently, the damping rate $\Gamma_{m}$ splits into two different values
$\Gamma_{m}^{+}$ and $\Gamma_{m}^{-}$, around $\Omega_{+}$ and $\Omega_{-}$,
respectively, as discussed in detail in Ref. \cite{Mickel}, where we also
observe that $\widetilde{\gamma}_{mn}\varpropto\left(  \Gamma_{m}^{+}%
-\Gamma_{m}^{-}\right)  /N$ while $\widetilde{\gamma}_{mm}\varpropto\left[
\Gamma_{m}^{+}+\left(  N-1\right)  \Gamma_{m}^{-}\right]  /N$. Therefore, for
Markovian white noise reservoirs, where the density of RM are the same around
both normal modes, such that $\Gamma_{m}^{+}\approx\Gamma_{m}^{-}$, the
cross-decay channels disappear together with the possibility of a DFS for
superposition states of the form $\left|  \Psi_{1,...,N}\right\rangle _{R=S}$
in both weak and strong coupling regime. In the weak coupling regime, we also
get the cross decay channels around zero even for reservoirs with spectral
densities distinct from the Markovian white noise, since in this case
$\Omega_{+}\approx$ $\Omega_{-}\approx\omega_{0}$ and again $\Gamma_{m}^{+}$
$\approx$ $\Gamma_{m}^{-}$. Therefore, differently from the former case of a
common reservoir, here the cross-decay channels linking the resonators to all
reservoirs besides their own, arise only from the strong coupling between the
resonators for non-Markovian white noise.

When distinct reservoirs are considered, the effectiveness of these
cross-decay channels depends on the condition $N\lambda/\omega_{0}\gtrsim1 $
(pulling the normal modes apart from each other) instead of the correlation
functions between the reservoir operators. Therefore, considering the strong
coupling regime and appropriate spectral density of the reservoirs, such that
$N\Gamma_{m}^{-}$ $\ll\Gamma_{m}^{+}$ and, consequently, $\widetilde{\gamma
}_{mn}\approx$ $\widetilde{\gamma}_{mm}$, the cross-decay channels\ become of
the same order of magnitude as direct-decay channels, inducing a DFS for
superpositions of the form $\left|  \Psi_{1,...,N}\right\rangle _{R=S}$. In
this case we get a master equation similar to Eq. (\ref{17}), apart from the
system-reservoir coupling strength $N\Gamma/2$, which becomes $\Gamma^{+}/2$
(assuming that all reservoirs have the same damping rate $\Gamma_{m}=\Gamma$).
(Note the lack of the factor $N$ in the latter coupling strength, following
from the dismissing of coupling around the normal mode $\Omega_{-}$.)
Therefore, for strongly coupled resonators, each one interacting with a
different reservoir, we get to same decoherence-free subspace states $\left|
\Psi_{1,...,N}\right\rangle _{R=S}$\ obtained in the case of a common
reservoir. We finally mention that the subject of distinct reservoirs have
analogies with previous work in literature \cite{Wu}, where external control
of quantum systems plays similar role as the strong coupling between distinct resonators.

\textbf{Acknowledgments}

We wish to express thanks to C. J. Villas-B\^{o}as and M. Fran\c{c}a for
helpful discussions, and for the support from FAPESP (under contract
\#02/02633-6) and CNPq (Intituto do Mil\^{e}nio de Informa\c{c}\~{a}o Qu\^{a}ntica).

\end{document}